\documentclass[12pt]{article}
\usepackage{graphicx}
\usepackage{amssymb}
\usepackage {amsmath}
\usepackage{latexsym}
\usepackage{epstopdf}

\begin{document}

\noindent
{\footnotesize{Hindawi Publishing Corporation
\textit{Research Letters in Physical Chemistry}
\textbf{Volume 2007}, Article ID 94286, 5 pages
doi:10.1155/2007/94286}}

\begin{center}
\section*{The study of influence of the Teslar technology on 
aqueous solution of  \\  some biomolecules}
\end{center}

\begin{center}
\bf {E. Andreev$^1$,   G. Dovbeshko$^1$ and V. Krasnoholovets$^2$}
\end{center}

\vspace{1mm}

\begin{center}
$^1$Department of Physics of Biological Systems, Institute of Physics, National Academy of Sciences, Prospekt Nauky 46, UA-03028 Kyiv, Ukraine \\
$^2$ Department of Physics, Institute for Basic Research, 90 East Winds Court, Palm Harbor, FL 34683, U.S.A.
\end{center}

\vspace{1mm}

\noindent
Correspondence should be addressed to V. Krasnoholovets, krasnoh@iop.kiev.ua \\
Received 5 June 2007; Accepted 25 July 2007  \\
Recommended by Leif A. Eriksson    \\

\begin{abstract}
The possibility of recording physical changes in aqueos solutions caused by a unique field generated by the Teslar chip (TC) inside a quartz wristwatch has been studied using holographic interferometry. We show that the refraction index of degassed pure distilled water and aqueous solutions of L-tyrosine and b-alanine affected by the TC does not change during the first 10 minutes of influence. In contrast, a 1\% aqueous solution of plasma extracted from the blood of a patient with heart-vascular disease changes the refractive index when affected by the TC. The characteristic time of reaction is about 10$^2$ s. Based on our prior research experience we state that the response of the system studied to the TC's field is similar to that stipulated by the action of a constant magnetic field with the intensity of $1.1 \times 10^{-3}$ T. Nevertheless, our team could unambiguously prove that the TC generates the inerton field, which is associated with a substructure of the matter waves (and, therefore, it does not relate to the electromagnetic nature).     

\bigskip

\noindent
\textbf{Key words:} Teslar chip, inerton field effects, holographic interferometer, reflective index, amino acid aqueous solution

\bigskip

\noindent
Copyright © 2007 E. Andreev et al. This is an open access article distributed under the Creative Commons Attribution License, which permits unrestricted use, distribution, and reproduction in any medium, provided the original work is properly cited.

\end{abstract}

\section*{\small  1. INTRODUCTION}

\noindent
         Numerous experiments fix the influence of electromagnetic field of certain frequency-amplitude ranges on living organisms. For instance, the magnetic field with frequencies in the range 0.3 to 30 Hz and with the intensity that is comparable with the Earth magnetic field can effectively influence the living organism function. It is supposed that the mechanism of influence should be connected with the parametric, or Schumann Resonance. The first four harmonics of the Schumann resonance are known: 7.8 Hz  $\pm$ 1.5 Hz, 14.5, 20, 26 Hz ($\pm$ 0.3 Hz) [1-3]. Well-known are two main mechanisms of the resonance reaction of the organism to a weak electromagnetic field. The first one is the Alfa-rhythm concerned with the thought process; the second one, the parametric resonance of organs, or organ systems, could be responsible for primary human reception [4-6]. A number of physiological processes, such as the reductive-oxidative process in living cells, responsible for the oxygen input, oxygen transport, etc. could be taken into account in this case. The parametric resonance of biological tissue and surrounding medium could be also responsible for the medical action of the TC.
         
        The aim of the present study is: 1) the influence of the TC on a biological model system and 2) registration of this influence in those cases when it is possible. The inventors [7] of this device state that the chip produces a longitudinal scalar wave/field (the notation was introduced by Nicola Tesla [8]). In this case a part of the energy is radiated in the form of a scalar longitudinal wave (also known as a Tesla free-standing wave). In more detail, this wave has been studied in Refs. [9,10].
        
         A model object must be sensitive; it has to have a large gain factor and the method must be reproducible and stable, simultaneously. Since our goal is to account for biophysical aspects of the influence of the TC on living organisms, the model system should include components available in hypodermic tissues of the wrist. These conditions allow us to choose, as the model of primary reception, the following: 
         
(i)	saturated aqueous solution of amino acids (tyrosine, tryptophane and alanine); 

(ii)	diluted aqueous solution of human blood plasma. (Although a solution of human serum albumin is a dominating in 
blood protein, the preparation of its solution by conventional method will mean that we obtain an equilibrium system, and it will be very difficult to move such system from its deep potential minimum. In the case of biomolecules of plasma of blood, which we study in this work, we deal with a non-equilibrium system and even very small stimuli applied to it could be effective.)

The parameter under study has become the refraction index $n$ of an aqueous solution.

\section*{\small 2. MATERIALS AND METHOD} 

          Holographic experiments have been carried out with the use of the holographic interferometer IGD-3, developed and produced in the Institute of Physics of Semiconductors of Nat. Acad. Sci. of Ukraine [11-13], whose optical scheme is given and described in Fig. 1. The He-Ne laser (1) radiation (power output equals 1 mW at   $\lambda$= 632.8 nm) is divided by the beam splitter cube (2) into two beams: the object beam and the reference beam. In the object beam shoulder there is the mirror (3) and the collimator (4) consisting of negative and positive lenses, which forms a parallel beam, 5 cm in diameter. The beam passes through the object under study (6), and then arrives at the finely dispersed diffuse scatterer (9). According to Lambert's law, its every point is scattering the light in all directions. The light from the whole surface of the scatterer arrives at every point of the light sensitive thermoplastic (10) in which plane the object can be selected. At one inclination of plate (5) the increase of n has to result in the increase of the interference period, i.e. in the decrease of the number of bands.
          
TC (7) has been put onto the top of a quartz cuvette (1 $\times$ 1 $\times$  4.3 cm$^3$) filled with the solution studied. If the dielectric characteristics of the object studied are the same before and after the TC influence, the fringe pattern remains unaltered and interference bands inside and outside the object's profile continue each other. On the contrary, if an external factor caused changes of n, the fringe pattern within the limits of the object's profile will change.          
         
         The amino acids used in our experiments were produced by Sigma, Inc. Aqueous solutions were prepared on the basis of pure bidistilled water. Prepared solutions, before the experiment, were maintained for 24 hours under 25 $^\circ$C. The plasma blood solution was extracted from the blood of a heart vascular disorder patient just after the blood was drawn at hospital by conventional methods. We dilute the solution by distilled water as 1: 50 and 1: 100. The time between the blood extraction and the holographic measurement was 4 hours.
        
        The procedure of dynamic measurement consisted of a sequence of records of interference patterns on a special thermoplastic plate, which then was fixed by a digital video camera. Afterward, the images were input into the computer and evaluated. For determination of the interference band centre, the 10 points along the horizontal line of cuvette have been chosen.
        
        The studies were conducted at temperature 20 $\pm$ 1.5 $^\circ$C   controlled by the thermocouple accurate to 0.2  $^\circ$C. We suppose that absolute meaning of the temperature does not influence the process under study due to the fact that we have been recording a dynamics of redistribution of the optical density. The most important point in the experiment was to protect the cuvette from the temperature gradient and the airflow. The last two disturbed factors have determined an inner non-stability of the system. 
        
        In Fig. 2a, typical interference patterns of the aerial ambient space (``Air") and the aqueous solution (``Solution") are presented. Vertical black lines show the image of the cuvette corner (its size 1 $\times$ 1 $\times$ 4 cm$^3$). Thus in our experiments we have been able to observe an alteration of the reflective index in the surface zone of the cuvette equal to 1$\times$ 2 cm$^2$ that is determined by the cuvette size and the aperture of laser beam.

\begin{figure}
\includegraphics[width=5in]{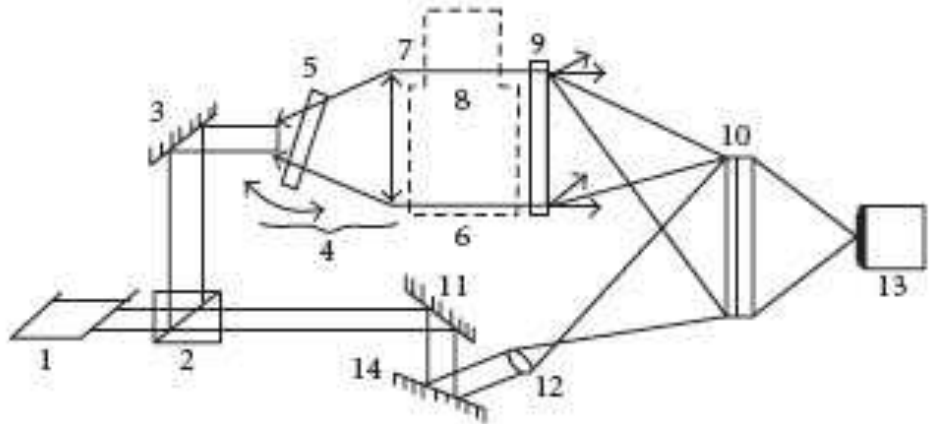}
\caption{\small Experimental holographic set. 1 Ð He-Ne laser; 2 Ð beam splitter cube; 3 Ð mirror; 4 Ð collimator; 5 Ð plane-parallel plate; 6 Ð quartz flask (cuvette) with the solution; 7 Ð Teslar bracelet; 8 Ð filter that divides two flasks; 9 Ð scattering layer; 10 Ð thermoplastic recording plate; 11 Ð reference beam mirror; 12 Ð reference beam lens; 13 Ð TV camera. }
\end{figure}

        In Fig. 2a, typical interference patterns of the aerial ambient space (``Air") and the aqueous solution (``Solution") are presented. Vertical black lines show the image of the cuvette corner (its size 1$\times$ 1 $\times$ 4 cm$^3$). We have been able to observe an alteration of $n$ in the surface zone 1 $\times$ 2 cm$^2$ of the cuvette (the cuvette size) and the aperture of laser beam. Deformations of the interference pattern in different points of the solution have been caused by changes in $n$ in these points. The resolution is defined by a location of the optical wedge, namely, by a sum of horizontal interference lines. The space resolution is about 2 mm. 

\begin{figure}
\includegraphics[width=3in]{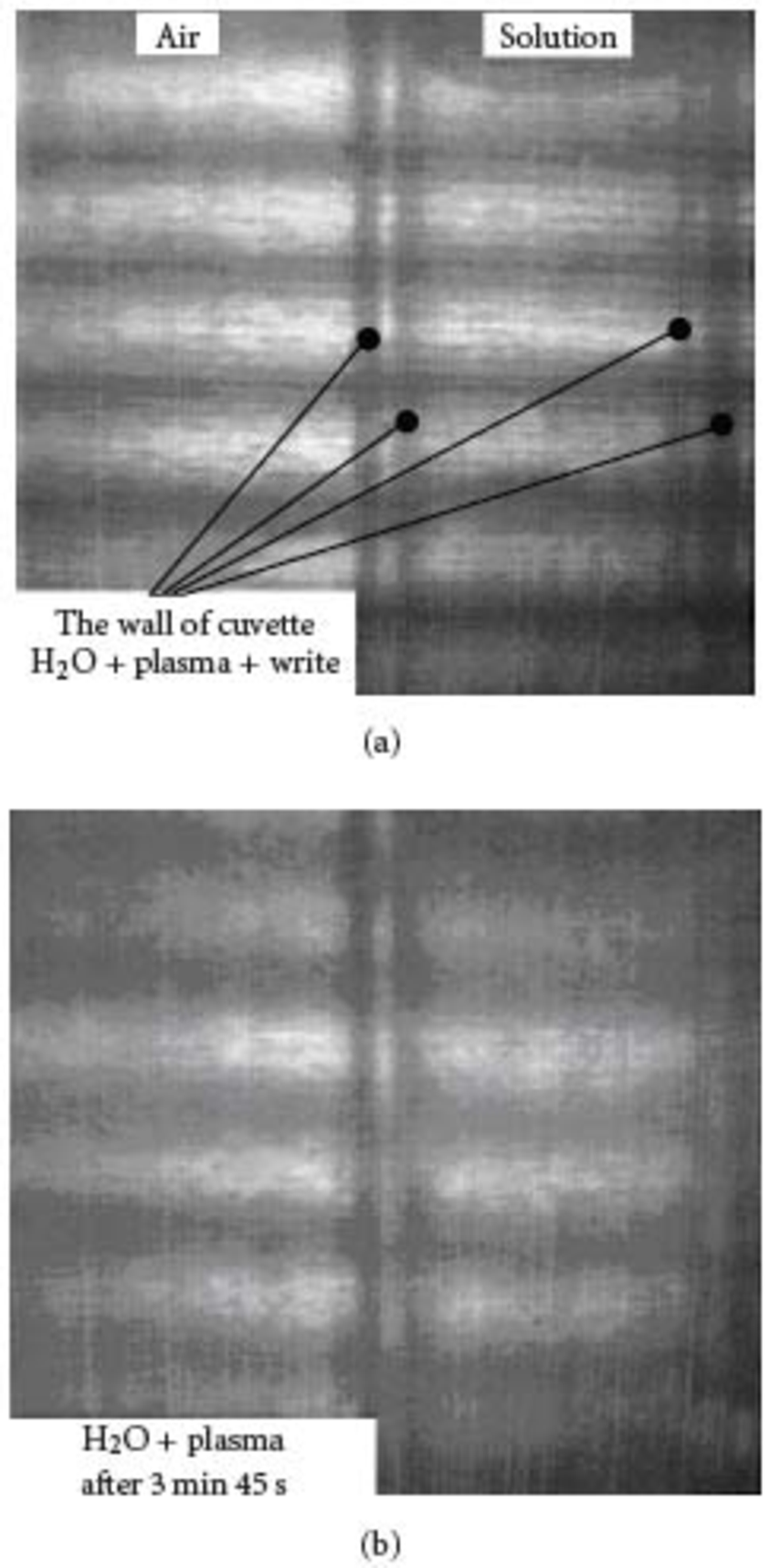}
\caption{\small Dynamics of the fringe pattern of the aqueous solution of plasma blood of human without the influence of the TC. The value of the effect is estimated by difference between the shift of the interference band in the cuvette with the solution and the position of same band in the air; we evaluated the shift of interference bands before and after the TC application. It is seen that during 4 minutes the bands in the cuvette have not been deformed and they have essentially not moved relative to those in the air. }
\end{figure}

  The method described gives a possibility to follow the response of the solution with the time factor of minimal discontinuous ability equal to 10 s. A sequence of pictures of the fringe pattern characterizes the space dynamics of the system studied in any place of the cuvette. As an example, Fig. 2b shows the fringe pattern formed in about 4 minutes starting from the moment of action of the TC that has been spaced at 2 mm from the cuvette. Changes of interference bands occurred during this time are associated with an internal stimulus. 

\section*{\small 3. RESULTS}
         
         A primary series of the experiments was conducted with distilled water, the saturated aqueous solution of L-tyrosine and $\beta$-alanine at 25  $^\circ$C . The experiments were conducted both in the morning and afternoon. The results showed typical slight changes of the fringe pattern in 400 s or larger time interval. These changes should be associated with the inner drift of liquid parameters. The curve of long-time dynamics does not show any influence on the side of the TC approximate to the cuvette.
         
         The other behavior and picture have been observed in the case of the blood plasma solution. Without the TC action this solution has shown stable and reproducible characteristics during more than 4 hours.    
         
         During more than one hour the system of recording and the objects of study (the solution of plasma and water) were stable and reproducible.
        
        In Fig. 3 we present the image of the cuvette with plasma blood solution affected by the TC. Black dots indicate the center position of one of the interference bands on the image plane. The value of the shift relating to zero line characterizes the degree of influence of the TC.  Black rectangles (Fig. 3, right) show the positions of two Teslar chips relating to cuvette;  the fringe pattern of the solution relating to the chip is deformed in different ways in different zones (short, mid, and far-distance). A physical mechanism of the change is associated with the increase of n in the short-distance zone; $n$ remains unchanged in the mid-distance zone and decreases in the far-distance zone.
        
        Moreover, it seems that slow laminar flows have been induced by the TC near the front wall and directed to it.

 \begin{figure}
\includegraphics[width=3in]{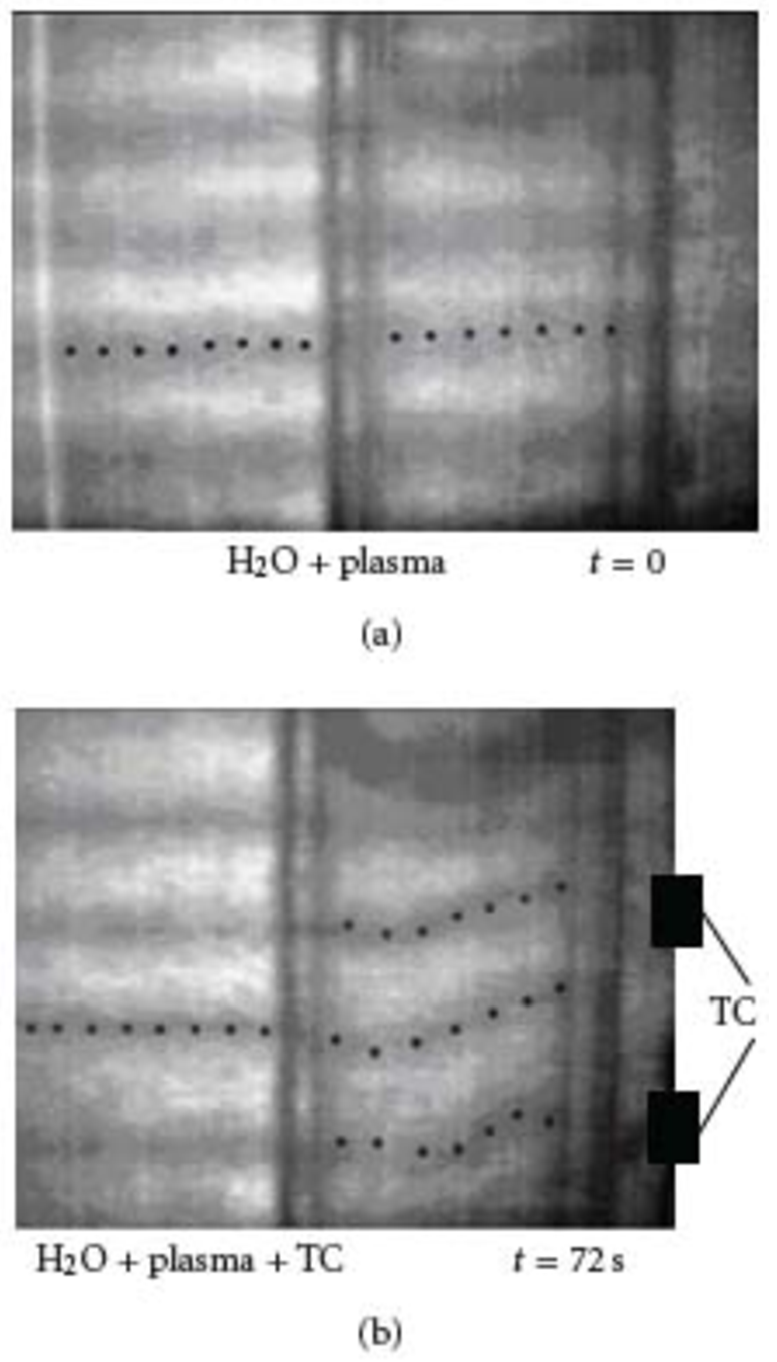}
\caption{\small Dynamics of the fringe pattern of the aqueous solution of plasma of human blood after the insertion of 2 TC. The strong disturbance of the optical density of the solution is emerged already in 72 s, right figure.  (The back covers of two sections of the bracelet are found at 4 mm from the right wall of the cuvette). }
\end{figure}

If the refraction index of the solution changes in one place under the influence of an external factor, the length of optical path will also change. With the purpose of the registration of the changes, the device is designed in such a way that the ``starting interferogram" constitutes a family of horizontal bands, bands of equal thickness. Depending on the character of changes of the optical density in the cuvette volume, the bands can be distorted (local changes of n), gaps between bands can expand without deformations (volumetric decreases of $n$). Thus arbitrary deformations of the fringe pattern are caused by a combination of local and global changes of the optical density. 

        Changes of $n$ are produced by changes in the structure of the network of hydrogen bonds of water, which being under the influence of oxygen, biomolecules and the inerton field generated by the TC, forms long-lived structures. In the mentioned network those new structures try to minimize the total energy relative to the volume occupied by the water system. Such kinds of changes (structuring of the aqueous solution) occur sufficiently slowly and therefore allows the recording by optical methods.
        
        The strongest changes in effects associated with the TC have been detected during the first 5 to 15 minutes starting from the moment of influence. The changes in the fringe pattern have been irregular in time. The saturation effect reaches at 10 minutes. Without the TC the number of interference bands remains the same in the field of the object and around it and is equal to five. Changes in the refractive index $n$ of the sample affected by the TC estimated from equation 

\begin{equation}
L \Delta n = \lambda \Delta k,
\label{eq1}
\end{equation}
where $L$ is the thickness of the sample (the aqueous solution studied), $\Delta n$ is the change in refractive index,  $\lambda$ is the wavelength of the source of light (laser), $\Delta k$  is the change in the number of interference bands as a result of an external effect. 

Influence of the TC on water leads only to minor changes of the fringe pattern ($\Delta n = 2 \times 10^{-5}$). The behavior of proteins is mainly determined by the influence of the TC. Effects associated with the TC and heating effects have shown the opposite trend/tendency. The temperature rise in the flask detected by the thermo sensor with 1 mW/cm$^2$ power density ranged between 0.2 to 0.5 $^\circ$C. Therefore, heating caused by the laser radiation allows an evaluation of the role of temperature. The estimation of the temperature effect by using the thermal conductivity equation and the thermal balance equations show the following. The maximum heating of the aqueous solution without account of the thermal exchange, i.e. under the condition unfavorable for the thermal effect estimation, may amount to 1 $^\circ$C. The calculation shows that even without the heat exchange between the aqueous solution and the environment the radiation effect with the power density of 10 mW/cm$^2$ may produce an increase of 1 $^\circ$C of the temperature of the 1.5 cm$^3$ volume of aqueous solution within 6 minutes. The temperature coefficient of changes in $n$ of water makes up $\Delta n = 2 \times 10^{-5}$. 

         The maximum change of n of the protein solution affected by the TC reached the value of  $\Delta n = 2 \times 10^{-4}$, which is an order of magnitude larger than the temperature changes of $n$. Thus the numerical estimates and the experimental data show that changes of $n$ caused by the influence of the TC have been conditioned by non-thermal changes of the solution dielectric constant, which may be described as the total contribution of electronic, vibration and orientational components.
         
In Fig. 4, experimental dots show changes of $n$ of the solution [the vertical axis] at the cuvette's back wall against the upper TC (rectangles) and the lower TC (triangles); recall the two TC are located near the front wall (see Fig. 3). Current time, in seconds, is plotted along the horizontal axis. Legends ``H$_2$O" (the blue background) and ``H$_2$O + Plasma" (the orange background) indicate different solutions in the cuvette. Time intervals are singled out for: 1) the magnet (``Magnet" on the green background) is applied to the front wall of the cuvette and 2) two links of the bracelet with two TC are set into the cuvette (the violet background). Moments of intermix of the solution (``Destruction") are shown by means of arrows; the intermix was made by using a medical syringe with 0.2 mm-needle: the solution was absorbed from the cuvette by the syringe and then poured back.

\begin{figure}
\includegraphics[width=5.5in]{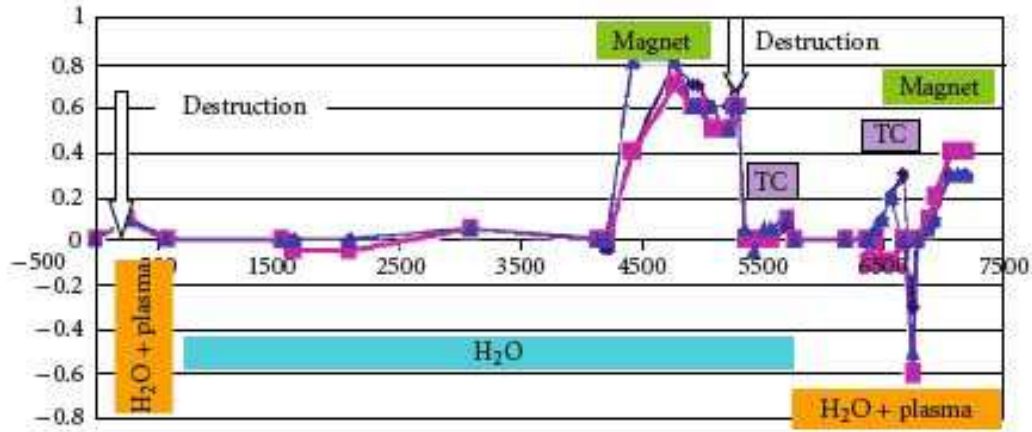}
\caption{\small Dynamics of changes of the fringe pattern of the cuvette's volume at different external factors. The horizontal axis shows current time in seconds.}
\end{figure}

\section*{\small 4. CONCLUSION }

        The TC does not affect distilled water. However, biomolecules of plasma of blood or an ensemble of such biomolecules in a micromolar concentration in water lead to the changes in reological characteristics, which allows the observation by optical methods, in particular, by the holographic interferometer. In the aqueous solution of blood plasma, biomolecules play a role of primary receptors of the TC radiation. 
        
Changes in the aqueous solution affected by the TC cover all the macroscopic volume of the sample studied. This behavior can be associated with both inner convective flows (like in the case of Benar cells) and structural changes of water. The latter may bring about changes in the reflective index of the solution and the fringe pattern.
Comparative responses of the aqueous solution to the mechanical, magnetic and TC influences point to a very specific action of the latter. A microscopic physical consideration of the phenomenon of the TC has already been performed in some detail [9,10,14]. We could prove [9,10] that the TeslarÕs phenomenon belongs to the inerton field effects and hence does not relate to the electromagnetic nature. The inerton field (the field of inertia) appears as a basic field in the submicroscopic mechanics of canonical particles developed in the real physical space and accounts for the availability of the wave $\psi-$function in conventional quantum mechanics (see, e.g. Refs. [15,16]). This field transfers local deformations of space, which appear in physical terms as mass. Therefore, the inerton field transfers mass changing potential properties of the environment. 

Consequently, the defect of mass  $\Delta m$ becomes an inherent property not only of atomic nuclei but any physical and physical chemical systems [14] (including biophysical ones). A more extensive study and repeated examinations should be completed to shed more light upon the mechanism of action of the Teslar chip and the inerton field in general upon living organisms.

\end{document}